\RequirePackage{fix-cm}
\documentclass{svjour3}                     
\smartqed  
\usepackage{graphicx}
\journalname{Orig Life Evol Biosph}
\begin{document}

\title{Silica aerogel for capturing intact interplanetary dust particles for the Tanpopo experiment
}

\titlerunning{Silica aerogel for capturing intact interplanetary dust particles}        

\author{Makoto Tabata \and
	Hajime Yano \and \\
	Hideyuki Kawai \and
	Eiichi Imai \and \\
	Yuko Kawaguchi \and
	Hirofumi Hashimoto \and
       Akihiko Yamagishi 
}


\institute{M. Tabata \at
              Department of Physics, Chiba University, Chiba 263-8522, Japan\\Institute of Space and Astronautical Science (ISAS), Japan Aerospace Exploration Agency (JAXA), Sagamihara 252-5210, Japan\\
              \email{makoto@hepburn.s.chiba-u.ac.jp}           
           \and
           H. Yano \and Y. Kawaguchi \and H. Hashimoto \at
              Institute of Space and Astronautical Science (ISAS), Japan Aerospace Exploration Agency (JAXA), Sagamihara 252-5210, Japan
	    \and
           H. Kawai \at
              Department of Physics, Chiba University, Chiba 263-8522, Japan
	    \and
           E. Imai \at
              Department of Bioengineering, Nagaoka University of Technology, Nagaoka 940-2188, Japan
	    \and
           A. Yamagishi \at
              Department of Applied Life Sciences, Tokyo University of Pharmacy and Life Sciences, Hachioji 192-0392, Japan
}

\date{Received: 30 October 2014 / Accepted: 28 November 2014}

\maketitle

\begin{abstract}
In this paper, we report the progress in developing a silica-aerogel-based cosmic dust capture panel for use in the Tanpopo experiment on the International Space Station (ISS). Previous studies revealed that ultralow-density silica aerogel tiles comprising two layers with densities of 0.01 and 0.03 g/cm$^3$ developed using our production technique were suitable for achieving the scientific objectives of the astrobiological mission. A special density configuration (i.e., box framing) aerogel with a holder was designed to construct the capture panels. Qualification tests for an engineering model of the capture panel as an instrument aboard the ISS were successful. Sixty box-framing aerogel tiles were manufactured in a contamination-controlled environment.
\end{abstract}

\section{Introduction}
\label{sec:1}

The Tanpopo mission is an astrobiological experiment to be performed on the Japanese Experiment Module (JEM) of the International Space Station (ISS) (Yamagishi et al. 2007 \cite{Yamagishi2007}, 2009 \cite{Yamagishi2009}). The primary goal of the mission is to test whether terrestrial microbes with dust ejected by events such as volcanic eruptions can reach the Earth's lower orbit and to analyze interplanetary dust particles containing prebiotic organic compounds, which could migrate among solar system objects. To achieve the nearly intact capture of cosmic dust, including terrestrial dust, interplanetary dust, and artificial space debris with a diameter of tens of micrometers and an orbital velocity of several kilometers per second, we developed an ultralow-density silica aerogel (Tabata et al. 2005 \cite{Tabata2005}, 2010 \cite{Tabata2010}, 2011 \cite{Tabata2011}, 2014 \cite{Tabata2014}). Silica aerogel, an amorphous, highly porous solid made of silicon dioxide, is a highly suitable medium for cosmic dust sampling because it is lightweight and optically transparent (Tsou 1995 \cite{Tsou1995}; Burchell et al. 2006 \cite{Burchell2006}; Brownlee et al. 2006 \cite{Brownlee2006}). After a sampling period of one year outside the JEM on orbit, the returned samples will be biochemically analyzed in ground laboratories using state-of-the-art methods. To protect the fragile aerogel, we designed a special holder, which we call the capture panel (Tabata et al. 2014 \cite{Tabata2014}). Inside the JEM, capture panels with twelve aerogel tiles will be attached to three different faces of the box-shaped Exposed Experiment Handrail Attachment Mechanism (ExHAM), newly developed by the Japan Aerospace Exploration Agency (JAXA). The ExHAM is attached to the Exposed Facility of the JEM exterior using a robotic arm and airlock. The exposure of the aerogel capture media for one year will be repeated three times by replacing the capture panels with aerogels.

\section{Overview of silica aerogel development and recent update}
\label{sec:2}


Our approach in developing silica aerogel as a cosmic dust capture medium is based on a method of producing hydrophobic, low-density, and transparent silica aerogel established in the 1990s by Japan's High Energy Accelerator Research Organization (KEK) in collaboration with Matsushita Electric Works Co., Ltd., Japan (currently Panasonic Corporation) (Adachi et al. 1995 \cite{Adachi1995}). For example, 2 m$^3$ aerogel tiles with densities varying from 0.04 to 0.11 g/cm$^3$ were fabricated using the classic KEK method and used as Cherenkov radiators in charged-particle identification devices in a high-energy physics experiment at KEK (Sumiyoshi et al. 1998 \cite{Sumiyoshi1998}). With the same method, aerogel blocks with a density of 0.03 g/cm$^3$ were manufactured by Matsushita Electric Works and were then used for capturing cosmic dust in the Earth's lower orbit in the Micro-Particles Capturer (MPAC) experiment of JAXA aboard the ISS (e.g., Noguchi et al. 2011 \cite{Noguchi2011}). In the above applications, originally hydrophilic aerogels were rendered hydrophobic to resist the age-related degradation owing to moisture absorption.

The low-density aerogel fabrication procedure of the classic KEK method consists of three steps (see Tabata et al. 2012 \cite{Tabata2012}). First, wet silica gel is synthesized using an appropriate mold (e.g., polystyrene case) and the sol--gel method. The primary chemical reactions are hydrolysis, condensation, and polymerization between a type of siloxane and water in ethanol with aqueous ammonia solution as a catalyst. We can adjust the density of the produced aerogels by varying the mixing ratio of the raw chemicals. After aging the wet gel to strengthen the silica three-dimensional network (e.g., for one week), the wet gel is extracted from the mold and subjected to hydrophobic treatment in an ethanol bath; the treatment was originally reported by Yokogawa and Yokoyama (1995) \cite{Yokogawa1995} and updated by Tabata et al. (2012) \cite{Tabata2012}. Finally, the wet gel is placed in an autoclave filled with ethanol and dried using a solvent extraction apparatus and supercritical carbon dioxide or ethanol. Before the final step, the impurities in the wet gel generated in the previous steps were removed by immersing new ethanol three times in the bath. It took approximately one month to complete all the fabrication steps.

In 2005, we reported the first successful production of \textit{hydrophobic} silica aerogels with a density of approximately 0.01 g/cm$^3$ (Tabata et al. 2005 \cite{Tabata2005}). We began developing ultralow-density aerogels back in 2003. At that time, we considered that the most significant problem was the shrinkage of the wet gels owing to the high temperature in the supercritical drying step. Ultralow-density aerogels have high porosity; therefore, we could decrease the air-filled pores and minimize the aerogel volume (i.e., density increase). We attempted to prevent the aerogel shrinkage by synthesizing an ultralow-density wet silica-gel layer (below 0.02 g/cm$^3$) on a rather high-density wet silica-gel layer (e.g., 0.04 g/cm$^3$). During the wet-gel synthesis, the stacked two layers are chemically combined to create a monolithic aerogel comprising two layers with different densities. The high-density base layer resists thermal shrinkage, and the attached upper layer is an ultralow-density aerogel. The technique for producing ultralow-density aerogels was eventually replaced by another method; however, we developed a method for producing multilayer aerogels with different densities.

In 2007, we succeeded in developing hydrophobic silica aerogel tiles of monolithic single layers with 0.01 g/cm$^3$ density by extending the classic KEK technique (Tabata et al. 2010 \cite{Tabata2010}). We adjusted the mixing ratio of raw chemicals to achieve ultralow density with no layer stacking. The production procedure was described in Tabata et al. (2011) \cite{Tabata2011}. Wet gels were synthesized using a polystyrene mold and carefully handled to prevent cracking. The typical dimension of the obtained aerogel tiles was 10 cm $\times $ 10 cm $\times $ 2 cm.

We then established a fabrication procedure to prevent aerogels from being contaminated by microbial DNA. Most of the aerogel fabrication processes were performed in a clean booth (class 1000) dedicated to producing aerogels for use in the Tanpopo mission to avoid possible contamination by dust particles in air. Washable tools were washed using a detergent (5\% Extran MA01, Merck, Germany) and ultrapure water. The background contamination level of the microbial DNA in the experimentally fabricated aerogels with approximately 0.01 g/cm$^3$ density was investigated using the polymerase chain reaction method. The level was below the detection limit; therefore, we concluded that the fabricated aerogels could be used in the Tanpopo experiment, at least from the perspective of contamination in the production process (Tabata et al. 2011 \cite{Tabata2011}).

The use of two-layer aerogels with different densities had been planned for the Tanpopo experiment (Yamagishi et al. 2007 \cite{Yamagishi2007}, 2009 \cite{Yamagishi2009}). A monolithic aerogel tile with an area less than 10 $\times $ 10 cm$^2$ comprises the top and base layers with densities of 0.01 and 0.03 g/cm$^3$, respectively. The lower density top layer will minimize the initial impact shock between the aerogel surface and the hypervelocity cosmic dust grains. On the other hand, the low-density aerogel requires large thickness to stop the dust grains. Consequently, we introduced the 0.03 g/cm$^3$ aerogel layer at the downstream side of the 0.01 g/cm$^3$ aerogel layer to immobilize the high-energy dust grains inside the aerogel tile. Each thickness of the top and base layers was set at approximately 1 cm at this stage. The 0.03 g/cm$^3$ base layer also protects the 0.01 g/cm$^3$ top layer from mechanical damages.

Using a two-stage light-gas gun (LGG), laboratory experiments simulating the hypervelocity capture of cosmic dust with ultralow-density aerogels were performed. The LGG enables us to fire a projectile into the target aerogel at a maximum velocity of 6--7 km/s. Particles 30--100 $\mu $m in diameter simulating the cosmic dust grains were enclosed in a dividable hollow bullet as the projectile, and the bullet was fired into the aerogel that was placed in a vacuum chamber. The bullet was centrifugally divided in flight and stopped by the bullet stopper in front of the target chamber. Thus, only the particles impact the aerogel. Tabata et al. (2011) \cite{Tabata2011} used soda-lime glass microspheres with a nominal diameter of 30 $\mu $m as the projectile and the experimentally fabricated two-layer aerogel tiles with a total thickness of approximately 2 cm as the target. The glass beads were fired at a velocity of approximately 6 km/s and were successfully captured inside the base layer of the two-layer aerogel. Kawaguchi et al. (2014) \cite{Kawaguchi2014} discussed a potential method for detecting microbes in the captured dust grains using fluorescence microscopy after staining around the impact cavities in an aerogel with a DNA-specific fluorescence dye. In that study, smectite clay particles with a diameter of 40--80 $\mu $m mixed with the \textit{Deinococcus radiodurans} bacterium were shot at approximately 4 km/s into the 0.01 g/cm$^3$ aerogel. In addition, Murchison meteorite particles with a diameter of 30--100 $\mu $m were shot into another 0.01 g/cm$^3$ aerogel at approximately 4 km/s, and the particles extracted were analyzed using infrared and Raman spectroscopy. The experiment was successful, and the preliminary results were presented in Ogata et al. (2013) \cite{Ogata2013}.

%
\begin{figure*}
 \includegraphics[width=0.75\textwidth]{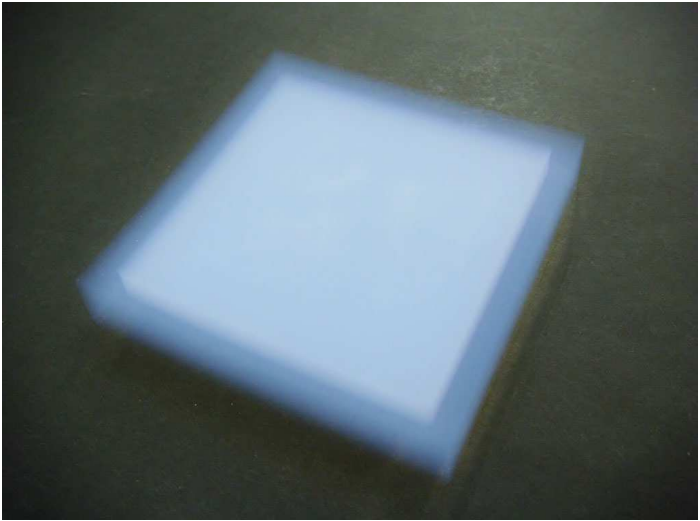}
\caption{Box-framing aerogel with a dimension of approximately 93 mm $\times $ 93 mm $\times $ 17 mm (sample that is not contamination controlled; ID = TNPP s1-2b). The 0.03 g/cm$^3$ layer, except the top surface, surrounds the 0.01 g/cm$^3$ layer.}
\label{fig:1}       
\end{figure*}

In 2012--2013, to protect the brittle low-density aerogel from damage caused by the rocket launch-induced vibrations, a capture panel with a dimension of 100 mm $\times $ 100 mm $\times $ 20 mm  made of aluminum alloy was designed (Tabata et al. 2014 \cite{Tabata2014}). In addition, to adopt the capture panel, a box-framing aerogel was designed and produced (Fig. \ref{fig:1}). The box-framing configuration consists of a 0.01 g/cm$^3$ top layer integrated in a 0.03 g/cm$^3$ hollow box-shaped base layer. This enables us to fix the aerogel tile to the capture panel by compressing only the 0.03 g/cm$^3$ base layer, which is sufficiently strong to resist the vibration, i.e., the 0.01 g/cm$^3$ top layer never touches the open-top capture panel. Qualification tests, including launch vibration, and depressurization and repressurization tests for an engineering model of the capture panel (i.e., prototype box-framing aerogels and their holders) as an instrument aboard the ISS were successful (Tabata et al. 2014 \cite{Tabata2014}).

By the end of 2013, the remaining qualification tests for the capture panel were successfully completed, and the box-framing aerogels were mass produced. Space vehicle landing and splashdown shock tests were successful. The capture panel also passed the thermal cycle tests from $-$107$^\circ$C to 300$^\circ$C (reported by Mita H, Fukuoka Institute of Technology, Japan). We then manufactured 60 contamination-controlled box-framing aerogel tiles. In addition, several box-framing samples were fabricated in an environment that was not contamination controlled (Fig. \ref{fig:1}), for reference.



\section{Conclusion}
\label{sec:3}

The development progress of the silica-aerogel-based cosmic dust capture panel for use in the Tanpopo mission was reviewed, and the recent updates were presented. We discussed the potential performance of the ultralow-density aerogel as a cosmic dust capture medium. An engineering model of the capture panel, including the box-framing aerogel with densities of 0.01 and 0.03 g/cm$^3$, was developed. The qualification tests were completed, and the contamination-controlled box-framing aerogel was successfully mass produced.

\begin{acknowledgements}
The authors are grateful to the members of the Tanpopo Collaboration for their assistance. This study was supported by the Space Plasma Laboratory at the Institute of Space and Astronautical Science (ISAS), JAXA.
\end{acknowledgements}



\end{document}